\documentstyle[amssymb,12pt]{article}

\newtheorem{th}{Theorem}[section]
\newtheorem{lm}[th]{Lemma}
\newtheorem{co}[th]{Corollary}

\newcommand{\ov}{\overline}
\newcommand{\bra}[1]{\left | #1 \right \rangle}

\newcommand{\cc}{\mbox{$\Bbb C$}} 

\date{}

\begin{document}

\title{\bf Spatially Correlated Qubit Errors and\\
Burst--Correcting Quantum Codes}
\author{ F. Vatan\thanks{e--mail: {\tt vatan@ee.ucla.edu}} \hspace*{1cm} 
V. P. Roychowdhury\thanks{e--mail: {\tt vwani@ee.ucla.edu}}
\hspace*{1cm} M. P. Anantram\thanks{e--mail: {\tt anant@ee.ucla.edu}} \\
Electrical Engineering Department\\
UCLA\\
Los Angeles, CA 90095}

\maketitle

\begin{abstract}
We explore the design of quantum error-correcting codes for cases
where the decoherence events of qubits are correlated.  In particular,
we consider the case where only spatially contiguous qubits decohere,
which is analogous to the case of burst errors in classical coding
theory. We present several different efficient schemes for
constructing families of such codes.  For example, one can find
one--dimensional quantum codes of length $n=13$ and 15 that correct burst
errors of width $b\leq 3$; as a comparison, a random-error correcting
quantum code that corrects $t=3$ errors must have length $n\geq 17$. In
general, we show that it is possible to build quantum
burst--correcting codes that have near optimal dimension. For example,
we show that for any constant $b$, there exist $b$--burst--correcting
quantum codes with length $n$, and dimension $k=n-\log n -O(b)$; as a
comparison, the Hamming bound for the case with $t$ (constant) random
errors yields $k\leq n - t \log n - O(1)$.

\end{abstract}

\section{Introduction}
The theory of quantum error correction is now an active area of
research \cite{Shor,calderbank,stean,Laflamme,bdsw,Gotesmann}.
Various techniques regarding construction of quantum error correcting
codes based on classical codes have been proposed
\cite{Shor,calderbank,stean,Gotesmann}.
These papers typically assume the `independent qubit decoherence'
(IQD) model; the decoherence of each qubit is uncorrelated to the
decoherence of any other qubit as they all interact with separate
reservoirs.

The IQD model is a valid assumption when the spatial separation of
qubits in a quantum register is larger than the correlation length of
the reservoir (or source of decoherence). Whether this condition is
met or not will depend on specific physical models for quantum
computers. The two main hardware proposals for quantum computers are
the ion trap model \cite{Cirac} and the polymer chain model
\cite{Lloyd}. The exact nature of decoherence in these models is not
well understood but we would expect the IQD model to break down in the
polymer chain model where qubits are only a few Angstroms apart.
Interaction with phonons whose spatial extent would be several atoms
long will result in correlated decoherence of spatially continuos
qubits.  Ref. \cite{Palma} provided the first step in studying the
effect of decoherence by assuming different models for interaction of
qubits and the reservoir. Specifically, the effect of decoherence on a
two-qubit system under circumstances when the IQD model is valid and
when the correlation length of the reservoir is larger than the
separation of the two qubits were studied. An important conclusion was
that the second model for decoherence (i.e., where the correlation
length is larger than the separation of qubits) leads to
superdecoherence and subdecoherence of certain off diagonal elements
of the density matrix in comparison to the IQD model.

If the details of the decoherence mechanism of qubits are known, then
it might be possible to build more efficient error correcting codes
compared to the IQD model. In classical error correction, it is well
known that magnetic tapes used for storage are usually defective over
length scales corresponding to a few bits.  Then it is sufficient to
code information on the tape so as to correct only spatially continuos
errors. Similarly, in a classical communication channel, disturbances
of the channel over short time periods lead to random single errors
while disturbances over longer time periods lead to temporally
continuos errors at the receiver.  Such errors are called burst errors
and the corresponding error correcting codes have significantly higher
rates (see, e.g., \cite{peterson}). 

In this paper, we discuss a quantum analog of burst-error correcting
codes. These codes would be important (i) when the coherence length of
the reservoir is large enough to cause decoherence of spatially
contiguous bits to be dominant, (ii) in storing of quantum information
on a string of qubits (this case is similar to the magnetic tape case
mentioned above; uninteded impurities here may perturb the energy
levels of a few contiguous qubits) and (iii) in communication of
quantum information where entangled qubits would be temporally
transported over an appropriate communication channel
\cite{bdsw}. We show in this paper that when such correlated errors 
are dominant, then codes with higher rates than those obtained using
the IQD model can be constructed.

The paper is organized as follows. In Section \ref{qcodes}, we give a
review of quantum error--correcting codes. Here we present revised
version of known schemes for constructing quantum codes such that they
can be applied for burst--error cases. In Section \ref{cycliccodes},
we give some results on binary cyclic codes, because these codes are
the backbone of our construction of quantum burst--correcting codes.
In Section \ref{explicitconstruction}, we present some explicit
constructions for quantum codes. In particular, we show how to
construct quantum codes that maps $k=n-2\log n -O(b)$ qubits to $n$
qubits that corrects bursts of width $b$. In Section
\ref{bounds}, we first derive the Hamming and Gilbert--Varshamov type bounds
for the maximum dimension of burst--correcting quantum codes. Then we show,
for any constant $b$, there exist $b$--burst--correcting quantum codes that
have near optimal dimension; i.e., they map $k$ qubits to $n$ qubits
where $k=n-\log n -O(b)$.

\section{Quatum error--correcting codes}
\label{qcodes}

In this section we provide basic definitions and notations about
quantum error--correcting codes. We shall also describe various
methods for constructing these codes; these techniques will be used in
Sections \ref{explicitconstruction} for constructing several different
kinds of burst-correcting quantum codes.

A sequence of amplitude errors in qubits $i_1,\ldots ,i_t$ of a block
of $n$ qubits can be represented by the unitary operator $X_\alpha$,
where the binary vector $\alpha$ of length $n$ has 1 components only
at positions $i_1,\ldots , i_t$. Thus, for the basis $\bra{v_1},\ldots
,\bra{v_{2^n}}$ of the $2^n$--dimensional Hilbert space of $n$ qubits,
we have
\begin{equation}
X_\alpha \bra{v_i}=\bra{v_i+\alpha} .
\label{aformula}
\end{equation}
Similarly, a sequence of phase errors can be written as 
\begin{equation}
Z_\beta \bra{v_i}=(-1)^{v_i\cdot \beta}\bra{v_i} ,
\label{pformula}
\end{equation}
where the binary vector $\beta$ represents the positions of errors,
and $v_i\cdot \beta$ is the inner product of two binary vectors modulo
2.  Note that
\begin{equation}
Z_\beta X_\alpha =(-1)^{\alpha\cdot \beta} X_\alpha Z_\beta . 
\label{apformula}
\end{equation}

Since we will be concerned with sets of errors with special
structures, it is useful for us to consider a general setting, where a
set ${\cal E}$ of possible errors of the form $\pm X_\alpha Z_\beta$
is fixed (a similar approach is followed in \cite{knilllaf}).  Let
$\overline{\cal E}$ be the set of the pairs $(\alpha ,\beta )$ such
that either $X_\alpha Z_\beta$ or $-X_\alpha Z_\beta$ is in $\cal E$.
For example, the result of the entanglment of introducing at most $t$
{\it random} errors in a state $\bra{x}$ can be repesented as $\pm
X_\alpha Z_\beta \bra{x}$, where $\mbox{wt}(\alpha \cup \beta )\leq
t$ \footnote{Here $\mbox{wt}(c)$ denotes the weight of the binary
vector $c$, i.e. the number of 1--components of $c$; and the binary
vector $\alpha
\cup \beta$ is the result of component-wise OR operation of $\alpha$
and $\beta$, for example $(011010)\cup (000110)=(011110)$.}. Therefore,
in this case $\ov{\cal E}=\bigl \{\, (\alpha ,\beta) \mid
\mbox{wt}(\alpha \cup \beta )\leq t \, \bigr \}$. 
We will use the following notations:
\begin{eqnarray*}
\ov{\cal E}_X & = & \left \{ \, \alpha \in \{ 0,1\}^n \mid
                    (\alpha ,\beta )\in \ov{\cal 
                    E}\ \mbox{for some}\ \beta \in \{ 0,1\}^n \, \right \}
                    \cup \{ \mbox{\bf 0}\}  , \\
\ov{\cal E}_Z & = & \left \{ \, \beta \in \{ 0,1\}^n \mid 
                    (\alpha ,\beta )\in \ov{\cal 
                    E}\ \mbox{for some}\ \alpha \in \{ 0,1\}^n \, \right \} 
                    \cup \{ \mbox{\bf 0}\} .
\end{eqnarray*}
For example, in the above example where $\cal E$ is the set of at most $t$
(random) errors, both $\ov{\cal E}_X$ and $\ov{\cal E}_Z$ are equal to
$\{ 
\, c\in \{ 0,1\}^n \mid \mbox{wt}(c)\leq t \, \}$. 
The following result gives a necessary and sufficient condition for a set 
of quantum states to constitute a quantum code.

\begin{th} {\em (\cite{bdsw}, \cite{knilllaf})}
A $2^k$--dimensional subspace $\cal Q$ of $\cc ^{2^n}$ is an $((n,2^k))$
error--correcting
quantum code mapping $k$ qubits to $n$ qubits that protect against all
errors in $\cal E$ if for every orthonormal basis $\bra{x_1},\ldots ,
\bra{x_{2^k}}$ of $\cal Q$ and every $e,e'\in {\cal E}$
\begin{eqnarray}
\left \langle x_i \, | \, ee'\, |\, x_j\right\rangle& = & 0, \qquad \hbox{if}\
i\neq j , \label{baab} \\
\left \langle x_i \, | \, ee'\, |\, x_i\right\rangle  & = &
\left \langle x_j \, | \, ee'\, |\, x_j \right\rangle . \label{baab2} 
\label{condition}
\end{eqnarray}
If for all $i$, $\left \langle x_i \, | \, ee'\, |\, x_i\right\rangle =0$,
then the quantum code is said to be {\em non-degenerate}.
\end{th}

In \cite{calderbank} and \cite{stean} it is shown how to use classical
error--correcting codes to build quantum codes. Although they stated
their results when errors are (random) errors of weight at most $t$,
their construction can easily be generalized for any set $\cal E$ of
errors. Before we state this result, we reiterate a definition
concerning classical codes. Let $\cal C$ be a subspace of $\{
0,1\}^n$, and $\cal F$ be a subset of $\{ 0,1\}^n$. We say $\cal C$
has the ability to correct every error from $\cal F$ (or simply, $\cal
C$ {\em has $\cal F$--correcting ability}) if and only if every two
different elements $e_1$ and $e_2$ of $\cal F$ belong to different
cosets of $\cal C$; or equivalently, $e_1+e_2\not \in {\cal C}$.

\begin{th}
Let ${\cal E}$ be a set of possible
quantum errors. If there are $[n,k]$ classical codes ${\cal C}_1$
and ${\cal C}_2$ (with $2k >n$) such that ${{\cal C}_2}^\perp \subseteq 
{\cal C}_1$ and ${\cal C}_1$ has $\ov{\cal E}_X$--correcting ability and 
${\cal C}_2$ has $\ov{\cal E}_Z$--correcting ability, then there is an 
$((n,2^{2k-n}))$ quantum code that has $\cal E$--correcting ability.
\label{construction}
\end{th}
{\bf Proof.} Consider cosets of ${\cal C}_1$ in ${{\cal C}_2}^\perp$;
i.e., the sets of the form $a+{{\cal C}_2}^\perp$ with $a\in {\cal
C}_1$. There are $2^{2k-n}$ different such cosets. Let $R=\{
a_1,\ldots,a_{2^{2k-n}}\}$ be the set of the representatives of
distinct cosets. For the basis of the quantum code, we consider
$2^{2k-n}$ vectors of the form $\displaystyle \sum_{c\in{{\cal
C}_2}^\perp}\bra{c+a}$, where $a\in R$.

To check that the condition (\ref{baab}) holds, consider vectors
\[ \bra{x}=\sum_{c\in{{\cal C}_2}^\perp}\bra{c+a}\qquad \mbox{and}\qquad 
 \bra{x'}=\sum_{c\in{{\cal C}_2}^\perp}\bra{c+a'} . \] (To simplify
the notation, throughout this paper we delete the normalization
factors.) If $a\neq a'$ then $\langle x \mid x'\rangle =0$.  Suppose
$e=(-1)^\lambda X_\alpha Z_\beta$ and $e'=(-1)^{\lambda '} X_{\alpha
'} Z_{\beta '}$ are in $\cal E$.  Note that
\begin{equation}
 \left \langle x \, | \, ee' \, |\, x'
\right\rangle =\epsilon \langle x\, |\, Z_{\beta+\beta'} 
X_{\alpha+\alpha'}\, |\,
x'\rangle ,
\label{xbaab}
\end{equation}
for some $\epsilon\in \{ -1, +1\}$. First suppose $\alpha +\alpha '\neq 
\mbox{\bf 0}$. Then the right hand side of (\ref{xbaab}) is equal to
\begin{equation}
 \epsilon \left \langle \sum_{c\in {{\cal C}_2}^\perp}\bra{c+a} \left | 
\sum_{c'\in{{\cal C}_2}^\perp} b_{c'} \bra{c'+a'+\alpha +\alpha '}\right \rangle
\right . , 
\label{inner}
\end{equation}
for some $b_{c'}\in \{ -1, +1\}$. If $c+a=c'+a'+\alpha +\alpha '$, for
some $c,c'\in {{\cal C}_2}^\perp$ and $a,a'\in {\cal C}_1$, then
$\alpha +\alpha ' \in {\cal C}_1$, which contradicts the $\ov{\cal
E}_X$--correcting ability of ${\cal C}_1$. Therefore, $c+a\neq
c'+a'+\alpha +\alpha '$, for every $c,c'\in {{\cal C}_2}^\perp$ and
$a,a'\in {\cal C}_1$, thus inner product (\ref{inner}) is equal to
zero. Now, suppose $\alpha +\alpha '=\mbox{\bf 0}$. If $a\neq a'$ then
then inner product (\ref{inner}) is obviously equal to zero. Finally,
suppose $\alpha +\alpha '= {\bf 0}$ and $a=a'$ (and of course $\beta
\neq \beta'$), then the right hand side of (\ref{xbaab}) is equal to
\[ \epsilon \sum_{c\in{{\cal C}_2}^\perp}(-1)^{(c+a)\cdot (\beta +\beta')} =
\epsilon (-1)^{a\cdot (\beta +\beta ')}\sum_{c\in {{\cal C}_2}^\perp} (-1)^
{c\cdot (\beta +\beta ')} . \] Now the sum in the right hand side of
the above equality is zero, because by $\ov{\cal E}_Z$--correcting
ability of ${\cal C}_2$ we have $\beta +\beta '\not \in {\cal C}_2$.
So the condition (\ref{baab2}) holds.  \hfill $\Box$

\vspace{8mm}
Special case of the above theorem is when ${\cal C}_1={\cal C}_2$;
thus we have the following corollary.

\begin{co}
Let ${\cal E}$ be a set of possible quantum errors. If there is an
$[n,k]$ classical code $\cal C$ (with $2k >n$) such that $\cal C$ is
weakly self--dual (i.e., ${\cal C}^\perp \subseteq {\cal C}$) and
$\cal C$ has both $\ov{\cal E}_X$--correcting ability and $\ov{\cal
E}_Z$--correcting ability, then there is an $((n,2^{2k-n}))$ quantum code
that has $\cal E$--correcting ability.
\label{sdconstruction}
\end{co}

It is possible to generalize the above construction even for the case
when $\cal C$ is not weakly self--dual.

\begin{th}
Let ${\cal E}$ be a set of possible quantum errors. Suppose there is
an $[n,k]$ classical code $\cal C$ such that $\cal C$ has $\ov{\cal
E}_X$--correcting ability and ${\cal C}^\perp$ has $\ov{\cal
E}_Z$--correcting ability.  Let ${\cal D}=\{ e+e'\, |\,
e,e'\in\ov{\cal E}_X\}$.  Then there is a quantum code that maps
$n-k-\lceil \log_2|{\cal D}| \rceil$ qubits to $n$ qubits and has
$\cal E$--correcting ability.
\label{gconstruction}
\end{th}
{\bf Proof.} We shall prove this theorem by developing an algorithm to
construct a quantum code. The basis of the 
quantum code consists of vectors of the form $\displaystyle
\bra{x_i}=\sum_{c\in {\cal C}}\bra{c+a_i}$, 
for some binary vectors $a_i\in \{ 0,1\}^n$. We shall show that one
can choose $2^{n-k-\lceil \log_2|{\cal D}|
\rceil}$ such $a_i$'s for the quantum code.

Let $a_1=\mbox{\bf 0}$, so $\displaystyle\bra{x_1}=\sum_{c\in {\cal
C}}\bra{c}$.  Suppose $a_1,\ldots ,a_{\ell -1}$ have already been
chosen. Then $a_\ell$ is any vector of $\{ 0,1\}^n$ which is not of
the form $c+a_i+\alpha +\alpha '$, for $c\in {\cal C}$, $1\leq i\leq
\ell -1$, and $\alpha , \alpha '\in\ov{\cal E}_X$.  In this process it
is possible to choose $m$ vectors $a_1,\ldots , a_m$ if
\[ m\cdot 2^k \cdot | {\cal D} | \leq 2^n . \]
The proof that conditions (\ref{baab}) and (\ref{baab2}) hold
is similar to the proof of Theorem \ref{construction}. \hfill $\Box$

\vspace{8mm}
 
In \cite{orthogeo,Gotesmann} a general method for describing and
constructing quantum error--correcting codes is proposed. Consider
unitary operators $e_1=X_{\alpha_1}Z_{\beta_1},\ldots
,e_k=X_{\alpha_k}Z_{\beta_k}$, such that ${e_i}^2=I$ (the identity
operator) and $e_ie_j=e_je_i$, for all $i$ and $j$ (i.e., $\alpha_i
\cdot \beta_i=0$ and $\alpha _i\cdot\beta_j+\alpha_j\cdot \beta_i=0$,
where the inner products are modulo 2). Consider the $k\times (2n)$
matrix
\begin{equation}
H=\left ( \begin{array}{c|c}
     \alpha_1 & \beta _1 \\ \vdots & \vdots \\ \alpha_k & \beta_k
          \end{array} \right ) . 
\label{stabilizer}
\end{equation}
Suppose the matrix $H$ has full rank over GF(2). Then the set of the
vectors $\bra{x}$ in $\cc ^{2^n}$ such that $e_i\bra{x}=\bra{x}$, for
all $1\leq i\leq k$, form an $(n-k)$--dimensional quantum code. The
following theorem connects the error--correcting ability of this code
with the properties of the dual space of $H$ in $\{ 0,1\}^{2n}$.

\begin{th} {\em \cite{orthogeo}}
Let $\cal E$ be a set of quantum errors.  Suppose the $k\times (2n)$
matrix $H$ in {\em (\ref{stabilizer})} is totally singular, i.e.,
$\alpha_i \cdot \beta_i=0$ and $\alpha _i\cdot\beta_j+\alpha_j\cdot
\beta_i=0$ for all $i$ and $j$. Let $\cal C$ denote the $[2n,k]$ binary code
with $H$ as its generator matrix. Then the space of the vectors
$\bra{x}$ such that $X_{\alpha _i}Z_{\beta _i}\bra{x}=
\bra{x}$, for all $1\leq i\leq k$, is an $((n,2^{n-k}))$ quantum code
that has ${\cal E}$--correcting ability if for every $(\alpha_1 ,\beta_1),
(\alpha_2,\beta_2)\in\ov{\cal E}$ either $(\alpha_1+\alpha_2 ,\beta_1+\beta_2)
\in {\cal C}$ or $H\cdot (\beta_1+\beta_2 \mid \alpha_1+\alpha_2 )^T\neq 0$.
\label{stabilizertheo}
\end{th}

\section{Some results on binary cyclic codes}
\label{cycliccodes}

We construct different burst--correcting quantum codes (each with
different different dimensions) by utilizing Theorems
\ref{construction}, \ref{gconstruction} and \ref{stabilizertheo}
when the underlying classical codes are cyclic. In this section we
provide necessary facts and results concerning binary cyclic codes.

 A linear subspace $\cal C$ of $\{ 0,1\}^n$ is called a {\em cyclic
code} if $\cal C$ is closed under the cyclic shift operator, i.e.,
whenever $(c_0,c_1,\ldots ,c_{n-1})$ is in $\cal C$ then so is
$(c_{n-1},c_0,\ldots , c_{n-2})$. When dealing with cyclic codes, it
is much easier to identify each binary vector with a polynomial over
the binary field $F_2=\{ 0,1 \}$.  For this, we correspond to the
vector $c=(c_0,c_1,\ldots ,c_{n-1})$ in ${F_2}^n$ the polynomial
$c(x)=c_0+c_1x+\cdots +c_{n-1}x^{n-1}$ in $F_2[x]$. For example, the
vector $(1,0,0,1,1,0)$ corresponds to the polynomial $1+x^3+x^4$.

One of the basic properties of a cyclic code $\cal C$ is that $\cal C$
is generated by one of its codewords; in the sense that there is a
codeword in $\cal C$, represented by the polynomial $g(x)$, such that
every codeword $c(x)\in {\cal C}$ is a multiple of $g(x)$, i.e.,
$c(x)=q(x)\cdot g(x)$ for some polynomial $q(x)$. Here the identity
$c(x)=q(x)\cdot g(x)$ is hold in the quotient ring $F_2[x] / (x^n+1)$,
i.e., we identify $q(x)\cdot g(x)$ with $q(x)\cdot g(x) \bmod
(x^n+1)$. It is well--known that if the polynomial $g(x)$ generates
the cyclic code $\cal C$ of length $n$, then $g(x)$ is a factor of
$x^n+1$.  (For more details see, e.g., \cite{macwilliams}.) 

Some more useful notations:  The {\em reciprocal} of a polynomial
$f(x)=a_0+a_1x+\cdots +a_{m-1}x^{m-1}+a_mx^m$, with $a_m\neq 0$, is
$f^\star (x)=a_m+a_{m-1}x+\cdots +a_1x^{m-1}+a_0x^m$, which is
obtained from $f(x)$ by reversing the order of the coefficients. Then
$f^\star (x)= x^mf(x^{-1})$. The {\em exponent} of the polynomial
$f(x)\in F_2[x]$ is the least integer $s$ such that $f(x)$ divides
$x^s+1$.

We start with stating some easy facts about cyclic codes.

\begin{lm}
Let $\cal C$ be cyclic code of length $n$ generated by the polynomial
$g(x)=1+\alpha_1 x+\cdots +\alpha_{k-1}x^{k-1}+x^k$ in $F_2[x]$.

(a) If $w(x)=x^j+a_1x^{j+1}+
\cdots +a_{\ell -1}x^{j+\ell -1}+x^{j+\ell}$ is in $\cal C$, then $\ell \geq k$.

(b) If $w\in {\cal C}$, $w\neq 0$ and $w$ contains a block of $m$
consecutive 0's, then $m < n-k$.
\label{cyclm}
\end{lm}

{\bf Proof.} (a) Since $w(x)\in {\cal C}$ and $\cal C$ is cyclic, the
codeword represented by the polynomial $w'(x)=1+a_1x+\cdots +a_{\ell
-1}x^{\ell -1}+ x^\ell$ should be in $\cal C$ as well. This means
\[ w'(x)=\sum_{i=0}^{n-k-1}\beta_ix^ig(x) , \]
where $\beta _i\in \{ 0,1\}$ are unique and at least one of $\beta_i$
is nonzero (note that this equation holds in $F_2[x]$). The degree of
the right--hand side polynomial is at least $k$, so $\ell \geq k$.

(b) We can assume, w.l.o.g., that $w(x)=a_{m}x^{m}+\cdots
+a_{n-1}x^{n-1}$.  Then $\displaystyle w(x)=\sum_{i=0}^{n-k-1}
\alpha_ix^ig(x)$, for unique $\alpha _i\in \{ 0,1\}$.  It easily
follows that $\alpha_0=\cdots = \alpha _{m-1}=0$. The condition
$w(x)\neq 0$ holds only if $m < n-k$.
\hfill $\Box$

\vspace{8mm}
In the next theorem we formulate a necessary condition for a cyclic
code to be self--dual.

\begin{th}
Let a polynomial $g(x)$ of degree $k$ ($k\leq n/2$) generates a cyclic
code $\cal C$ of length $n$. Let $g(x) = g_1(x)\cdots g_m(x)$ be a
decomposition of $g(x)$ to irreducible polynomials. If the reciprocal
of any of $g_i(x)$ is not among $g_1(x),\ldots , g_m(x)$ (specially,
none of the $g_i(x)$ is self--reciprocal), then $\cal C$ is weakly
self--dual, i.e., ${\cal C}^\perp \subseteq {\cal C}$.
\label{selfdual}
\end{th}

{\bf Proof.} Suppose $x^n +1=g(x)h(x)$. The cyclic code ${\cal
C}^\perp$ is generated by the polynomial $h^\star (x)=x^{n-k}h({1\over
x})$. Let $g^\star (x)$ be the reciprocal of $g(x)$, i.e., $g^\star
(x)=x^kg({1\over x})$.  Then
\begin{eqnarray*}
  {\cal C}^\perp \subseteq {\cal C} & \mbox{iff} & h^\star (x)\in
{\cal C} \\ & \mbox{iff} & h^\star (x) =p(x)\cdot g(x)\ \mbox{for some
$p(x)$} \\ & \mbox{iff} & x^{n-k}\cdot {{1\over x^n}+1\over g({1\over
x})}=p(x)\cdot g(x) \\ & \mbox{iff} & x^n+1=p(x)\cdot g(x) \cdot
x^kg({1\over x}) \\ & \mbox{iff} & g(x)\cdot g^\star (x)\
\mbox{divides}\ x^n+1 .
\end{eqnarray*}
Let $g_i^\star (x)$ be the reciprocal of $g_i(x)$.  Then $g^\star
(x)=g_1^\star(x)\cdots g_m^\star(x)$. It is well--known (see, e.g.,
\cite{macwilliams}) that if $g_i(x)$ is not self--reciprocal, then
$g_i(x)\cdot g_i^\star (x)$ divides $x^n+1$. Therefore $g(x)\cdot
g^\star (x)$ divides $x^n +1$ and $\cal C$ is weakly self--dual.
\hfill $\Box$

\vspace{8mm}
Now we give some results on cyclic codes that correct burst errors.  A
{\em burst of width} $b$ is a vector in $\{0,1\}^n$ whose only nonzero
components are among $b$ successive components, the first and the last
of which are nonzero.  (The last component $c_{n-1}$ of the vector
$(c_0,c_1,\ldots ,c_{n-1})$ is understood to be adjacent to $c_0$.) As
mentioned in the previous section, we say a linear code $\cal C$ has
{\em burst--correcting ability} $b$ if for every bursts $w_1$ and
$w_2$ of width $\leq b$ we have $w_1+w_2\not
\in {\cal C}$. The following theorem gives a simple criterion for a cyclic
code to have burst--correcting ability.

\begin{th}
Let $\cal C$ be a cyclic code generated by the polynomial $g(x)$ of
degree $k$. If $k\geq {n \over 2}+b$, then $\cal C$ has
burst--correcting ability $b$.
\label{burstdeg}
\end{th}

{\bf Proof:} Suppose $w_1$ and $w_2$ are bursts of width $\leq b$.
Then in $w_1+w_2$ there is a block of at least ${n\over 2}-b$
consecutive 0's.  Since ${n\over 2}-b \geq n-k$, by Lemma \ref{cyclm}
(b), the theorem follows. \hfill $\Box$

\vspace{8mm}
The following theorem by Fire \cite{fire} and Melas and Gorog
\cite{melas} (see also \cite{peterson}) gives a general method to
construct interesting burst--correcting cyclic codes.

\begin{th} Let $q(x)$ generate an $[n',k']$ code that has burst--correcting
ability $b$. Let $p(x)$ be an irreducible polynomial of degree $\geq
b$ and exponent $e$ such that $(p(x),q(x))=1$ (i.e., $p(x)$ and $q(x)$
have no common factor). Then the cyclic code $\cal C$ of length
$n=en'$ generated by $c(x)=q(x)p(x)$ has burst--correcting ability
$b$.
\label{fire}
\end{th}

In the following theorem we construct a small burst--correcting code.
The interesting property of this code is that it is {\em weakly
self--dual}; the property which is not addressed by the previous
theorem.

\begin{th} The polynomial $g(x)=1+x+x^2+x^4+x^5+x^8+x^9$ generates a cyclic
$[21,12]$ code ${\cal C}$ which has burst--correcting ability $b=4$.
Moreover, $\cal C$ is weakly self--dual.
\label{n21}
\end{th}

{\bf Proof.} First note that $g(x)$ is factored to irreducible
polynomials as
\[ g(x)=(1+x+x^3)(1+x^2+x^4+x^5+x^6) .\]
It is easy to see that the factors of $g(x)$ are factors of $x^{21}+1$ (or
look at the table of factors of $x^n+1$ in \cite{macwilliams}). So $g(x)$ 
generates a [21,12] cyclic code. From Theorem \ref{selfdual} it follows that
$\cal C$ is weakly self--dual.

It can be shown that $\cal C$ has burst--correcting ability $b=4$.
\hfill $\Box$

\vspace{8mm}
To utilize Theorem \ref{fire} for producing cyclic weakly self--dual
codes that correct $b>4$ bursts, we need to start with small cyclic
weakly self-dual codes with burst--correcting ability $b$. It seems it
is hard to find such codes with optimal, or near optimal, length. But
it is possible to construct small cyclic weakly self--dual codes that
correct {\em random} $t$ errors. In the next lemma we give a
construction for such codes. Although in this way we do not get
optimal codes, the result is enough to rise to an infinite class of
burst--correcting codes.

\begin{lm}
For any $t$, there is a binary cyclic weakly self--dual $[n,k,2t+1]$
code with $n=2^m-1$ and $k=n-tm$, where $n > 2(2t-1)^2$.
\label{bchcode}
\end{lm}

{\bf Proof.} The code is a BCH code. We describe it by its cyclotomic
cosets mod $n$ (for details see \cite{macwilliams}). The cyclotomic
cosets which define the code are $C_{2i-1}$, for $1\leq i\leq t$.
Since $2t-1<\sqrt{n}$, all these cosets are distinct (see p.262 of
\cite{macwilliams}). So the code has minimum distance $2t+1$. To prove
the code is self--dual, we have to show for every $i$ and $j$, $1\leq
i,j\leq t$, $n-(2j-1)\not\in C_{2i-1}$. If this is not true, then
$n-(2j-1)=(2i-1)2^\ell$, for some integer $\ell$. This implies that
$2^\ell$ should divide $2j$. But by assumption on $n$, we have $2^\ell
> 2t$; which ends up in a contradiction.  \hfill $\Box$

\section{Explicit construction of burst--correcting quantum codes}
\label{explicitconstruction}

The {\em quantum burst--correcting codes} are defined naturally. Consider
the set $\cal E$ of quantum errors such that both $\ov{{\cal E}_X}$ and
$\ov{{\cal E}_Z}$ are bursts of width $\leq b$. Then any quantum code that
has $\cal E$--correcting ability is called a $b$--burst--correcting code.

First we show that there is a two--dimensional quantum code of length
$n=15$ which corrects burst errors of width $b=3$. From the table
given in \cite{gf4}, it follows that to correct $t=3$ (random) errors
one qubit should be mapped to at least 17 qubits.

We consider the cyclic [15,9] code ${\cal C}$ generated by
$1+x^3+x^4+x^5+x^6$. As it is noted in \cite{peterson}, this code
corrects $b=3$ burst errors.  We show that the dual of this code has
the same burst--correcting ability.

The dual code ${\cal C}^\perp$ is generated by the polynomial 
$1+x+x^4+x^5+x^6+x^9$. So the following is a generator matrix for 
${\cal C}^\perp$:
\[ \left [ \begin{array}{ccccccccccccccc}
   1 & 1 & 0 & 0 & 1 & 1 & 1 & 0 & 0 & 1 & 0 & 0 & 0 & 0 & 0 \\ 
   0 & 1 & 1 & 0 & 0 & 1 & 1 & 1 & 0 & 0 & 1 & 0 & 0 & 0 & 0 \\ 
   0 & 0 & 1 & 1 & 0 & 0 & 1 & 1 & 1 & 0 & 0 & 1 & 0 & 0 & 0 \\
   0 & 0 & 0 & 1 & 1 & 0 & 0 & 1 & 1 & 1 & 0 & 0 & 1 & 0 & 0 \\ 
   0 & 0 & 0 & 0 & 1 & 1 & 0 & 0 & 1 & 1 & 1 & 0 & 0 & 1 & 0 \\ 
   0 & 0 & 0 & 0 & 0 & 1 & 1 & 0 & 0 & 1 & 1 & 1 & 0 & 0 & 1
   \end{array} \right ] . \] 

We want to show that if $b_1$ and $b_2$ are two burst of width $\leq
3$ and $b_1\neq b_2$, then $b_1+b_2\not \in {\cal C}^\perp$. First
note that $b_1+b_2$ contains a block of at least 5 consecutive 0's.
Then w.l.o.g. we can assume $b_1+b_2=(000000\star \star \cdots \star)$
or $b_1+b_2=(000001\star
\star \cdots
\star)$. If $b_1+b_2\in {\cal C}^\perp$, then
in the first case we would have $b_1+b_2=0$, i.e., $b_1=b_2$ which
contradicts the assumption $b_1\neq b_2$; and in the second case
$b_1+b_2=(000001100111001)$ which is not sum of two bursts of width
$\leq 3$.  This completes the proof that ${\cal C}^\perp$ corrects
bursts of width 3.

Now we show that the all--one vector {\bf 1} is not in coset of any
vector of the form $b_1+b_2$, where $b_1$ and $b_2$ are bursts of
width $\leq 3$. Assume, by contradiction, that $\mbox{\bf
1}+b_1+b_2\in {\cal C}^\perp$. Since $b_1+b_2$ has a block of at least
5 consecutive 0's, $\mbox{\bf 1}+b_1+b_2$ is either $(111110\star
\star \cdots \star)$ or $(111111\star \star \cdots \star)$. In the
first case $\mbox{\bf 1}+b_1+b_2$ is $(111110111010001)$ and in the
second case it is $(111111011101000)$. So $b_1+b_2$ is either
$(000001000101110)$ or $(000000100010111)$; which in neither case it
can be the sum of two bursts of width $\leq 3$.

So the desired quantum code consists of $\displaystyle
\bra{0_L}=\sum_{c\in{\cal C}} \bra{c}$ and $\displaystyle
\bra{1_L}=\sum_{c\in{\cal C}} \bra{c+\mbox{\bf 1}}$.

Eric Rains \cite{rains} has brought to our attention that there is
a 3--burst--correcting quantum code with smaller length. This is a $[[13,1,5]]$
code. The stabilizer of this code is defined by a quadratic residue code over
$\mbox{GF(4)}=\{\, 0,1,\omega,\ov{\omega}=\omega ^2\, \}$ with 
$g(x)=(1+x)g_1(x)$ as its generator polynomial, where 
$g_1(x)=1+\ov{\omega}x+\omega x^3+\ov{\omega}x^5+x^6$
(see \cite{gf4} for details on quantum codes defined by codes over GF(4)).
Here it is enough to show that the cyclic code $\cal C$ (over GF(4)) generated
by the polynomial $g_1(x)$ is a 3--burst--correcting code. Suppose the nonzero
polynomial $q(x)$ (of degree $\leq 12$) represents 
a codeword in $\cal C$ that is a sum of two bursts of 
width $\leq 3$. Then at least 7 coefficients of $q(x)$ are zero. Therefore,
without loss of generality, we can assume $q(x)=1+a_1x+\cdots +a_8x^8$. Hence
$q(x)=(1+ax+bx^2)g_1(x)$, for some $a,b\in\mbox{GF(4)}$. But for any $a$
and $b$ the corresponding $q(x)$ is not a sum of two bursts of width $\leq 3$. 

Now we show the existence of infinite classes of quantum
burst--correcting codes.

\begin{th}
If there is a binary $[n,k]$ code $\cal C$ (with $k<{n\over 2}$) such
that $\cal C$ and ${\cal C}^\perp$ both have burst--correcting ability
$b$, then there is an $((n,2^K))$ quantum code with $K=n-k-2\lceil \log
n \rceil -b$ that corrects all burst errors of width $b$.
\label{qburst1}
\end{th}

{\bf Proof:} It follows from the fact that the number of vectors of
the form $w_1+w_2$, where $w_1$ and $w_2$ are bursts of width $\leq
b$, is at most $n^2\cdot 2^b$ and the Theorem \ref{gconstruction}.
\hfill $\Box$

\begin{co}
There are $((n,2^k))$ quantum codes with $n=(2^m-1)(2b-1)$ and
$k=n-m-2\lceil \log n \rceil -3b+1$ having burst--correcting ability
$b$.
\end{co}

{\bf Proof.} From Theorem \ref{fire} with $n'=2b-1$, 
$q(x)=x^{2b-1}-1$ and $p(x)$
any primitive polynomial of degree $m$ such that $m > 2b-1$, we get an
$[n,n-m-2b+1]$ binary code $\cal C$ having burst--correcting ability
$b$.  If $m\geq 3$, then we can apply Theorem~\ref{burstdeg} to show
that ${\cal C}^\perp$ also has burst--correcting ability $b$. Then the
result follows by applying Theorem~\ref{qburst1}. \hfill $\Box$

\vspace*{8mm} 
For fixed constant $b$, the above result gives a family of quantum
codes of length $n$ and dimension $n-3\log n-O(1)$ having
burst--correcting ability $b$.  In the next theorem we show for $b\leq
4$ we can construct burst--correcting quantum codes with dimension
$n-2\log n-O(1)$.

\begin{th}
For every $m\geq 7$, there is an $((n,2^k))$ quantum code with
$n=21(2^m-1)$ and $k=n-2m-18$ having burst--correcting ability $b=4$.
\label{n21m}
\end{th}

{\bf Proof.} We utilize Theorem \ref{fire} for $q(x)$ the degree 9
polynomial given in Theorem \ref{n21} and $p(x)$ any primitive 
polynomial of degree $\geq 7$. 
Since $p(x)$ is primitive, it is not self--reciprocal. Thus we obtain a
binary $[n=21(2^m-1),k=n-m-9]$ code ${\cal C}$, which is weakly
self--dual by Theorem \ref{selfdual}.  Now by applying Corollary
\ref{sdconstruction}, we get the quantum code with the given
parameters. \hfill $\Box$

\vspace{8mm}
By utilizing Lemma \ref{bchcode}, we can get a similar result for the
case $b>4$.  The following theorem shows how to construct $((n, 2^K))$
quantum codes, with $K= n-2\log n-O(b)$, for constant burst-width $b$; note
however that the constant, $O(b)$, in the preceding expression could 
be a large function of $b$. 

\begin{th}
For every $b$, there is an $((n,2^k))$ quantum $b$--burst--correcting code, 
where $n=(2^{m'}-1)(2^m-1)$, $k=n-2m-2bm'$, $2^{m'}>2(2b-1)^2$ and $m>m'$.
\end{th}

{\bf Proof.} As in the previous theorem, we utilize Theorem
\ref{fire}, where $q(x)$ is the generator polynomial of the BCH code of
length $n'=2^{m'}-1$ given in Lemma \ref{bchcode}, and $p(x)$ is any
primitive polynomial of degree $m>m'$.  \hfill $\Box$

\section{Bounds and a new scheme for construction}
\label{bounds}

In this section we present general upper and lower bounds for maximum
dimension of a quantum burst--correcting code.

\begin{th}
Let $\cal E$ be a set of errors, and ${\cal D}=\{ e+e'\, |\,
e,e'\in{\cal E}\}$.  Let $2^k$ be the maximum of dimension of any
non-degenerate quantum code of length $n$ that has $\cal
E$--correcting ability. Then
\[ n-\log_2|{\cal D}| \leq k \leq n-\log_2 |{\cal E}| . \]
\label{gbound}
\end{th}

{\bf Proof.} Let $\bra{x_1},\ldots ,\bra{x_{2^k}}$ be a basis for a
quantum code of dimension $2^k$. The upper bound {\em (The Hamming
bound)} follows from this fact that, by (\ref{baab}), in the space
$\cc ^{2^n}$ the $2^k |{\cal E}|$ vectors of the form $X_\alpha
Z_\beta \bra{x_i}$ should be orthogonal, and hence independent.

The lower bound {\em (The Gilbert--Varshamov bound)} follows by an
argument similar to the one given in \cite{orthogeo}.
\hfill $\Box$

\begin{co}
Let $2^k$ be the maximum of dimension of any non-degenerate quantum
code of length $n$ which has burst--correcting ability $b$. Then
\[   n-2 \log_2 n -2b \leq k \leq n- \log_2(n-b+2)-2b-3 . \]
\label{bbound}
\end{co}   
   
{\bf Proof.}
It is enough to note that in this case $|{\cal E}|=3\cdot 4^{b-1}(n-b+2)$,
and $|{\cal D}|\leq 4^bn^2$. 
\hfill $\Box$

\vspace{8mm}

Next we present a new scheme for constructing quantum codes from
classical linear codes. By utilizing this method, for fixed constant
$b$, we obtain $b$--bursts--correcting quantum codes of length $n$
with dimension $n-\log_2 n -O(1)$. These are almost optimal codes
(compare with the bound given in Corollary \ref{bbound}).

\begin{th}
If there is a $(3b+1)$--burst correcting binary $[n,k]$ cyclic code
$\cal C$ such that $\cal C$ is weakly self--dual, then there is a
$b$--burst--correcting $((n,2^k))$ quantum code.
\label{burstcyccons}
\end{th}

{\bf Proof.} Suppose the $(n-k)\times n$ matrix $H$ is a parity check matrix 
for $\cal C$. Let $H^{\rightarrow m}$ denote the matrix that is obtained from
$H$ by shifting (cyclically) the columns $m$ times to the right. Note
that $H^{\rightarrow m}$ is also a parity check matrix of $\cal C$, because
$\cal C$ is cyclic. Now, consider the stabilizer defined by the matrix
\[ G= \left [ H+H^{\rightarrow b}\, |\, H+H^{\rightarrow 2b+1}\right ] . \]
It is easy to check that $G$ is indeed a totally singular matrix.
Suppose $e=(e_1\mid e_2)$ and $e'=(e'_1\mid e'_2)$ are bursts of width $\leq
b$, and $e\neq e'$.  Let
\[ w=e_1+e'_1+(e_1+e'_1)^{\rightarrow b}+e_2+e'_2+
(e_2+e'_2)^{\rightarrow 2b+1},\] 
where $e^{\rightarrow b}$ denotes the
vector obtained by cyclically shifting $e$ to the right $b$ times.
Then it is easy to check that $w\neq 0$ and $w$ is the sum of two
bursts of width $\leq 3b+1$. So $w\not\in{\cal C}$ and
\[ G\cdot (e+e')^T=H\cdot w^T\neq 0. \]
Now the theorem follows from Theorem \ref{stabilizertheo}.  \hfill $\Box$

\vspace{8mm}
To apply the above theorem, we need weakly self-dual
$b$--burst--correcting binary codes with arbitrary length. For $b\leq
4$, Theorem \ref{n21} gives explicit construction of such codes. In
general, we can apply the following theorem.

\begin{th} {\em \cite{abdel}}
For every $b$ and for every square--free polynomial $e(x)$ of degree $b-1$
and of index $m_e$ such that $e(0)\neq 0$
and for every sufficiently large $m\equiv 0(\bmod\ m_e)$,
a primitive polynomial $p(x)$ of degree $m$ exists such that 
$e(x)p(x)$ generates  a $b$--burst--correcting code of length $n=2^m-1$ 
and dimension $k=n-m-b$.
\label{abdelth}
\end{th}

To get weakly self--dual codes, choose $e(x)$ to be any primitive
polynomial of degree $b-1$. Then $e(x)p(x)$ generates a weakly
self--dual $b$--burst--correcting code, because no primitive
polynomial is self--reciprocal. Thus we get the following bound for
burst--correcting quantum codes.

\begin{th}
For every $b$ and for sufficiently large $n=2^m-1$ (where $m\equiv
0(\bmod\ m_b)$ for some fixed integer $m_b$ depending only on $b$),
there are $b$--burst--correcting quantum codes of length $n$ and
dimension $n-m-3b+1$.
\end{th}

\section*{Concluding Remarks} 

We described different schemes for constructing quantum
burst--correcting codes. As expected, these classes of codes are more
efficient than codes that protect against random errors. More
specifically, to protect against burst errors of width $b$ (where $b$
is a fixed constant), it is enough to map $n-\log n -O(b)$ qubits to
$n$ qubits, while in the case of $t$ random errors at least $n-t\log
n$ qubits should be mapped to $n$ qubits (the best construction so far
maps $n-(t+1)\log n-O(1)$ qubits to $n$ qubits).
 
We present an explicit construction of almost optimal quantum code for
bursts of width $\leq 4$. It would be interesting to find similar
explicit constructions for bursts of larger width.  Also, our
constructions work for specific values of $n$, the length of the
quantum code. It remains open how to generalize these constructions
for other values of $n$. Finally, in Theorem
\ref{burstcyccons} we develop a novel method to obtain quantum codes from
binary codes such that the rate of the quantum code has the same order
of magnitude as the binary code. If this method can be generalized to
other classes of quantum codes, there would be a great improvement in
the rate of the existing quantum codes.

{\bf Acknowledgements.} We would like to thank Eric Rains for helpful comments
and the example of $[[13,1]]$ 3--burst--correcting code.

\end{document}